\documentclass[sigconf]{acmart}
\usepackage{tabularx}
\usepackage{makecell} % for thicker hlines
\usepackage{bchart}
\usepackage{pgfplots}
\usepackage{xcolor}
\definecolor{mintgreen}{RGB}{135,232,133}
\definecolor{darkblue}{RGB}{255,192,118}
\definecolor{orange}{RGB}{250,128,128}
\AtBeginDocument{%
  }

\setcopyright{acmcopyright}
\copyrightyear{2024}
\acmYear{2024}
\acmConference[BNU-HKBU-UIC '24]{Undergraduate Consortium}{Dec,
  2024}{Zhuhai, Guangdong}
\acmPrice{15.00}
\acmISBN{978-1-4503-XXXX-X/23/08}

\begin{document}

\title{Identification of Epileptic Spasms (ESES) Phases Using EEG Signals: A Vision Transformer Approach}

\author{Wei Gong}
\authornote{Both authors contributed equally to this research.}
\email{r130005022@mail.uic.edu.hk}
\affiliation{%
  \institution{Beijing Normal University-Hong kong Baptist University United International College}
  \city{ZhuHai}
  \state{Guangdong}
  \country{P.R.C.}
}

\author{Yaru Li}
\authornotemark[1]
\email{r130018031@mail.uic.edu.hk}
\affiliation{%
\institution{Beijing Normal University-Hong kong Baptist University United International College}
  \city{ZhuHai}
  \state{Guangdong}
  \country{P.R.C.}
}

\renewcommand{\shortauthors}{Weu Gong and Yaru Li}

\begin{abstract}
This work introduces a new approach to the Epileptic Spasms (ESES) detection based on the EEG signals using Vision Transformers (ViT). Classic ESES detection approaches have usually been performed with manual processing or conventional algorithms, suffering from poor sample sizes, single-channel-based analyses, and low generalization abilities. In contrast, the proposed ViT model overcomes these limitations by using the attention mechanism to focus on the important features in multi-channel EEG data, which is contributing to both better accuracy and efficiency. The model processes frequency-domain representations of EEG signals, such as spectrograms, as image data to capture long-range dependencies and complex patterns in the signal. The model demonstrates high performance with an accuracy of 97\% without requiring intensive data preprocessing, thus rendering it suitable for real-time clinical applications on a large scale. The method represents a significant development in the advancement of neurological disorders such as ESES in detection and analysis.
\end{abstract}


\maketitle

\section*{Introduction}

Electroencephalogram (EEG) signals are crucial in diagnosing and analyzing neurological disorders, such as Epileptic Spasms (ESES), which are characterized by continuous neuronal discharges during sleep. EEG data, often represented as time series signals, capture intricate patterns of brain activity. However, the complexity and high-dimensional nature of EEG data make it difficult to analyze effectively, requiring sophisticated models that can interpret these complex signals accurately. Traditionally, various machine learning techniques, such as Convolutional Neural Networks (CNN) and Long Short-Term Memory (LSTM) networks, have been employed for EEG classification tasks, including ESES detection. These methods, while effective, have certain limitations that hinder their performance in real-world applications.

CNNs, a popular method for processing spatial data, are well suited for image classification tasks. However, when applied to EEG data, CNNs are often limited by the inability to capture long-range temporal dependencies and the complex, multidimensional patterns inherent in the data. Additionally, CNNs typically require extensive data pre-processing, which can lead to a loss of important signal information. On the other hand, LSTMs, which excel at handling sequential data, can capture long-range dependencies in time-series data. However, they may struggle with high-dimensional EEG data due to the sequential nature of the model, which can limit their ability to efficiently process large-scale datasets.

To address these challenges, this paper proposes a novel approach to ESES detection by converting EEG data into image representations and using Vision Transformers (ViT) for classification. Vision Transformers have demonstrated remarkable success in handling complex image data by leveraging self-attention mechanisms, allowing the model to focus on relevant spatial and temporal parts of the signal, making it highly suitable for EEG data analysis \cite{dosovitskiy2020transformers}. In our method, we first convert the frequency-domain data into grayscale and RGB images. Transformation of EEG signals into images allows us to take advantage of the power of ViT, a model originally designed for image classification, to handle the high-dimensional data and capture the complex patterns present in EEG signals \cite{dosovitskiy2020transformers}. Unlike CNNs and LSTMs, ViTs excel at processing long-range dependencies and capturing intricate relationships within the data using self-attention mechanisms, enabling the model to focus on relevant parts of the signal and enhance classification accuracy.

The benefits of using ViT for EEG classification are many. First, the conversion of EEG data into grayscale and RGB images provides a more precise and efficient representation of the signal, reducing the need for extensive manual pre-processing. Second, ViT's ability to process long-range dependencies within the data allows for more accurate detection of ESES events. The self-attention mechanism in ViT enables the model to capture subtle, distributed patterns across time and frequency bands, which is critical for diagnosing conditions like ESES \cite{parmar2018image}. Finally, the ViT model's ability to handle large datasets efficiently and quickly makes it particularly well suited for real-time clinical applications, offering a promising approach to improving diagnostic accuracy in neurological disorders.

In summary, this work introduces a new approach to ESES detection that outperforms traditional methods by converting the frequency data into grayscale images and leveraging the power of Vision Transformers. By converting EEG data into image representations and utilizing ViT's advanced self-attention mechanism, we provide a more accurate, efficient, and scalable solution for EEG-based diagnosis, setting our model apart from existing methods in the field.

\section{Related Work}

\section*{Current Circumstance}

ESES (Encephalopathy with Status Epilepticus during Sleep) is characterized by continuous or near-continuous epileptiform activity during sleep, typically detected through both time-domain and frequency-domain EEG analysis. Various machine learning and signal processing techniques are currently being utilized to identify ESES patterns in EEG data. A summary of some of the more prominent algorithms and models in use today is as follows:

\subsection*{Classic Signal Processing and Machine Learning}

Before the advent of deep learning models, detection of ESES was primarily performed using classic signal processing techniques combined with machine learning classifiers. These methods include:

\begin{itemize}
    \item \textbf{Fourier Transform}: 
    The Fourier Transform is used to transform EEG signals into the frequency domain, allowing for analysis of specific frequency bands such as delta, theta, alpha, beta, and gamma waves \cite{fourier2009}.
    \item \textbf{Wavelet Transform}: 
    This method is typically employed to detect non-stationary signals, such as seizures and other epileptic activities, due to its ability to analyze both the time and frequency domains simultaneously \cite{daubechies1992}.
    \item \textbf{Spectrograms}: 
    Spectrograms transform EEG signals into time-frequency representations using techniques such as Short-Time Fourier Transform (STFT), which are then classified using machine learning classifiers such as Support Vector Machines (SVM), Random Forests, or k-Nearest Neighbors (K-NN) \cite{spectrogram2015}.
\end{itemize}

\subsection*{Machine Learning Classifiers}

Once features are extracted using these transformations, classifiers such as Support Vector Machines (SVM), Random Forests, K-Nearest Neighbors (K-NN), or Artificial Neural Networks (ANNs) are commonly employed for detection \cite{svm1995, randomforest2001, knn2001}.

\begin{figure}[t]
  \centering
  \includegraphics[width=\columnwidth]{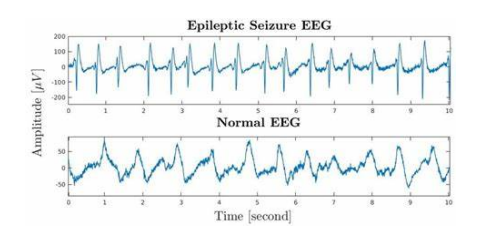}
  \caption{EEG normal vs ESES.png }
  \label{fig:EEG_signal}
\end{figure}

\begin{table}[htbp]
\centering
\begin{tabular}{|c|c|c|}
\hline
\textbf{Sample ID} & \textbf{EEG Data (16 channels, 500 Hz)} & \textbf{ESES Label} \\ \hline
1  & \([X_1, X_2, \dots, X_{500}]\) & 0 (N) \\ \hline
2  & \([X_1, X_2, \dots, X_{500}]\) & 1 (Y) \\ \hline
3  & \([X_1, X_2, \dots, X_{500}]\) & 0 (N) \\ \hline
4  & \([X_1, X_2, \dots, X_{500}]\) & 1 (Y) \\ \hline
5  & \([X_1, X_2, \dots, X_{500}]\) & 0 (N) \\ \hline
\vdots  & \vdots & \vdots \\ \hline
\end{tabular}
\caption{EEG data sampled at 500 Hz with corresponding ESES labels (0 = No, 1 = Yes).}
\label{table:eeg_data_labels}
\end{table}

\begin{figure*}[t]
  \centering
  \includegraphics[width=\textwidth]{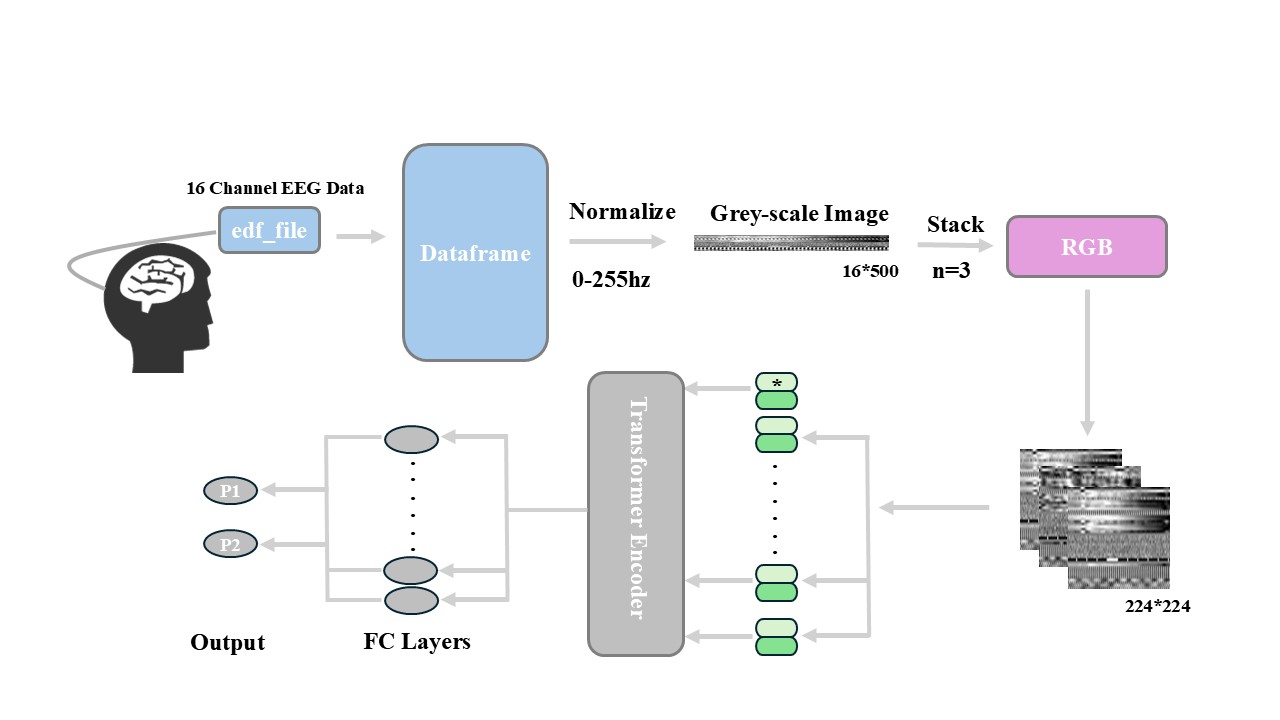}
  \caption{The Data Pre-processing and the Model Architecture}
  \label{fig:eegvit}
\end{figure*}

\section*{Signal-to-Image Preprocessing}

\begin{figure}[t]
  \centering
  \includegraphics[width=\columnwidth]{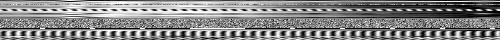}
  \caption{The Greyscale Image Formed By Original Signal }
  \label{fig:Greyscale}
\end{figure}

The EEG data used in this study originates from a dataset containing both EEG signals and other physiological data. For this task, we specifically focus on the EEG data, which consists of 16 channels, each corresponding to a different electrode on the scalp. These 16 channels of EEG data are sampled at a frequency of 500 Hz per second, resulting in a matrix of dimensions \(16 \times 500\), where each column represents a time-point and each row corresponds to one of the 16 channels.

Each EEG sample in the dataset is associated with a label that indicates the presence or absence of Encephalopathy with Status Epilepticus during Sleep (ESES). The label is binary, with a "yes" or "no" indicating whether the ESES event occurred during the second of the signal.

The preprocessing steps are as follows:
\begin{itemize}
    \item \textbf{Normalization}: The raw EEG data is first normalized to fit within the 0-255 range, which corresponds to the standard RGB color range. This is done to ensure that the data is suitable for image-based analysis, converting the signal values into a format that can be visualized as an image.
    
    \item \textbf{Greyscale Conversion}: The normalized data, which is initially in the form of a \(16 \times 500\) matrix, is transformed into a greyscale image. Since the values have been normalized to the 0-255 range, each value is mapped to a pixel intensity, resulting in a \(16 \times 255\) greyscale image. This step reduces the dimensionality of the data while preserving the important signal information.
    
    \item \textbf{RGB Stacking}: After the greyscale transformation, the image is duplicated across the three RGB channels to form a three-channel RGB image. This operation involves copying the same pixel intensity values into each of the three color channels (Red, Green, and Blue), resulting in a \(16 \times 255 \times 3\) RGB image. This allows the model to process the EEG data in a format compatible with vision-based models.
    
    \item \textbf{Resizing}: The RGB image is then resized to a standard image size of \(224 \times 224\) pixels. This resizing step ensures that all input images have the same dimensions, which is crucial for feeding the data into convolutional neural networks (CNNs) or other image-processing models. The resized image is now in the format \(224\times 224 \times 3\), ready for use as input to the model.
\end{itemize}

These preprocessing steps transform the raw time-domain EEG signals into a format suitable for image classification models, leveraging the power of deep learning techniques originally designed for visual data.

The processed data was then divided into training, validation in a 80\%, 20\%. The validation and test data sets were made up of distinct subjects. This approach allowed us to assess the model's ability to generalize and adapt to new data, providing a more robust and accurate evaluation of its potential real-world application.

\section{Model Architecture and Implementation}

After preprocessing the EEG data into \(224 \times 224\) RGB images, the data is fed into the Vision Transformer (ViT) model, specifically the \texttt{vit\_base\_patch16\_224\_in21k} architecture. This variant of ViT is designed for image inputs of \(224 \times 224\) resolution and leverages a patch-based mechanism to process images effectively. 

ViT divides the input image into patches of size \(16 \times 16\) and treats each patch as a "token," analogous to words in natural language processing. Each token is then passed through a transformer encoder that utilizes self-attention mechanisms to capture long-range dependencies across the entire image. The model is pre-trained on the ImageNet-21k dataset, providing a robust foundation for transfer learning to the domain-specific task of EEG classification \cite{dosovitskiy2020transformers}.

The self-attention mechanism in ViT allows the model to focus on relevant parts of the signal, making it particularly suited for complex datasets like EEG. By leveraging the power of transformers for image recognition, the ViT model captures intricate spatial relationships and long-range dependencies, enhancing its ability to classify the presence of ESES patterns accurately. 

The use of \texttt{vit\_base\_patch16\_224\_in21k} ensures compatibility with the input dimensions of the preprocessed EEG data and provides a scalable solution for handling high-dimensional datasets in neurological disorder detection.

\subsubsection{Vision-Transformer Backbone}

\begin{figure}[t]
  \centering
  \hspace*{-1cm}
  \includegraphics[width=\columnwidth]{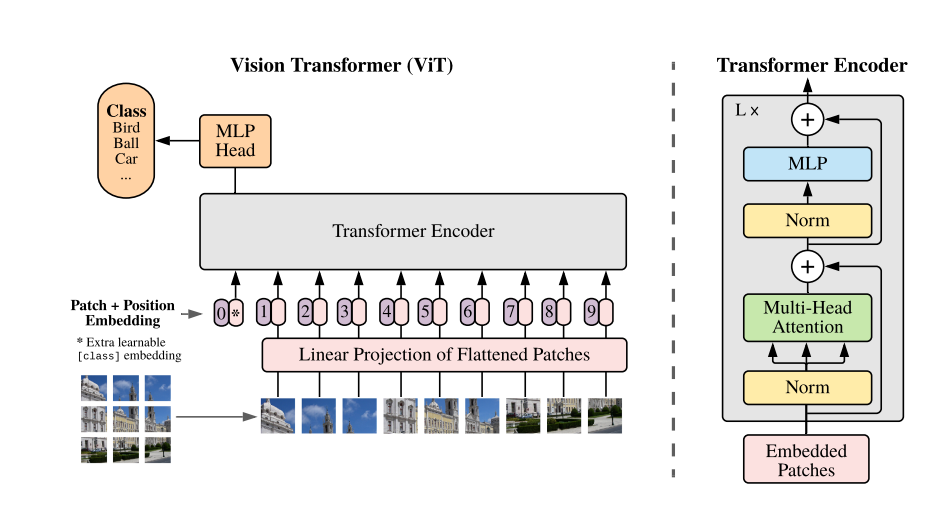}
  \caption{VIT Backbone by \cite{dosovitskiy2020transformers} }
  \label{fig:vit_block}
\end{figure}

\section{Training Configuration}

In this study, the model is trained using categorical data labels, where the presence of ESES is classified as either "yes" or "no." Given the categorical nature of the labels, we use Cross-Entropy as the loss function. Cross-Entropy is particularly suitable for multi-class classification tasks, as it measures the difference between the predicted probabilities and the actual class labels, enabling the model to learn effectively from the data.

The training is performed with a batch size of 64, which balances the trade-off between computational efficiency and model performance. This configuration ensures that the model is able to process a substantial amount of data at each iteration while maintaining an optimal update rate for the parameters.

The training is conducted on a Dell R730XD server equipped with a V100-PCIE-16GB GPU. The server is further supported by two E5-2690 V4 CPUs, which provide the necessary computational resources for handling large-scale datasets and model training. The choice of V100-PCIE-16GB GPU ensures high throughput during the training process, allowing the Vision Transformer (ViT) model to handle the large number of parameters efficiently.

The initial learning rate is set to 0.001, providing a stable starting point for model optimization. Additionally, the learning rate factor (lrf) is set to 0.01, which controls the rate of learning during training and helps achieve faster convergence while minimizing the risk of overshooting the optimal solution. These hyperparameters are fine-tuned to strike a balance between fast convergence and avoiding overfitting.

\begin{itemize}
    \item \textbf{Loss Function:} Cross-Entropy Loss
    \item \textbf{Batch Size:} 64
    \item \textbf{GPU:} V100-PCIE-16GB
    \item \textbf{CPU:} Dual E5-2690 V4
    \item \textbf{Initial Learning Rate:} 0.001
    \item \textbf{Learning Rate Factor (lrf):} 0.01
\end{itemize}

\section{Comparison with CNN Architecture}

To evaluate the performance of the Vision Transformer (ViT) model, we compared it with a standard Convolutional Neural Network (CNN) architecture. CNNs are widely used in image classification tasks due to their ability to extract spatial features through convolutional operations and pooling layers \cite{lecun1998gradient}. A typical CNN consists of multiple convolutional layers followed by non-linear activation functions, pooling layers for dimensionality reduction, and fully connected layers for classification.

For this comparison, we implemented a CNN architecture with several convolutional layers, ReLU activation functions, and max-pooling layers. The model was trained on the same preprocessed EEG dataset used for the ViT model, with the same training hyperparameters, including a batch size of 64, an initial learning rate of 0.001, and Cross-Entropy Loss as the objective function.

After 80 epochs of training on the dataset, the CNN achieved an accuracy of 94%, while the ViT model outperformed it with an accuracy of 97%. This result demonstrates the effectiveness of the self-attention mechanism in ViT for capturing long-range dependencies and complex patterns in EEG data compared to the localized feature extraction of CNNs.

\begin{table}[h]
\centering
\label{model_comparison}
\begin{tabular}{|c|c|c|}
\hline
\textbf{Model} & \textbf{Accuracy (\%)} \\ \hline
CNN & 94 \\ \hline
ViT (\texttt{vit\_base\_patch16\_224\_in21k}) & 97 \\ \hline
\end{tabular}
\caption{Comparison of CNN and ViT models on EEG dataset.}
\end{table}

\section{Advantages of Vision Transformer over CNN}

The experimental results highlight several key advantages of the Vision Transformer (ViT) model compared to the CNN architecture:

\begin{enumerate}
    \item \textbf{Global Feature Extraction:} Unlike CNNs, which primarily rely on local receptive fields for feature extraction, ViT utilizes a self-attention mechanism to model global dependencies in the input image. This allows ViT to better capture long-range relationships in EEG signal data.
    
    \item \textbf{Scalability with Data:} The transformer-based architecture of ViT can effectively scale with larger datasets, leveraging pretraining on extensive datasets to fine-tune performance on domain-specific tasks.

    \item \textbf{Performance Improvement:} After 80 epochs of training, ViT achieved an accuracy of 97\%, outperforming CNN's accuracy of 94\% on the same dataset. This demonstrates ViT's superior ability to generalize to complex patterns in ESES detection.

    \item \textbf{Flexibility in Input Representation:} The patch-based representation of input images in ViT provides greater flexibility for processing structured EEG data, as seen in this study.

    \item \textbf{Robustness to Noise:} ViT's attention mechanism enhances its robustness to noise, which is particularly beneficial for tasks like EEG analysis where noise is often present in the data.
\end{enumerate}

These findings underscore the potential of ViT as a powerful alternative to traditional CNNs for EEG-based medical diagnosis tasks.

\section{Future Deployment}
\begin{figure}[t]
  \centering
  \hspace*{-1cm}
  \includegraphics[width=0.8\columnwidth]{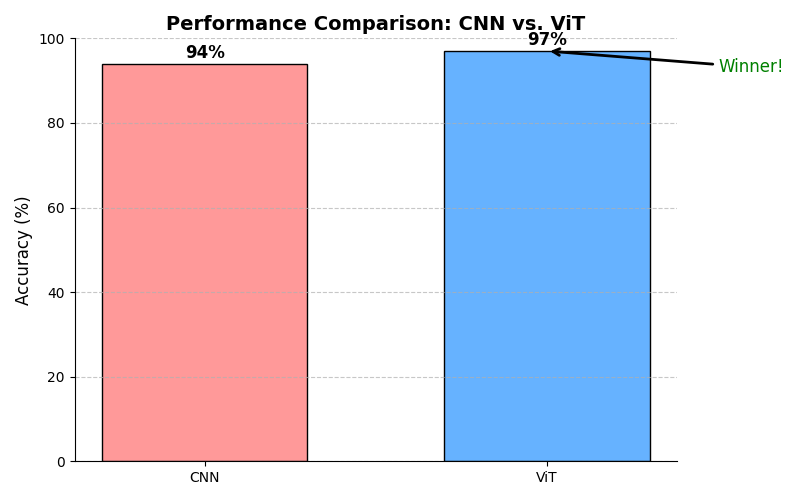}
  \caption{VIT vs CNN Architect }
  \label{fig:VIT-CNN}
\end{figure}

The Vision Transformer (ViT) architecture's inherent flexibility offers a significant advantage for future deployment, particularly in integrating multi-modal data. Unlike traditional convolutional neural networks (CNNs), which are primarily optimized for image-based tasks, ViT's transformer-based structure can seamlessly incorporate additional modalities such as text, metadata, or other sequential inputs. 

This capability is especially critical in the context of medical diagnosis, where combining EEG data with clinical notes, patient history, or other textual data could lead to more accurate and comprehensive predictive models. By leveraging the self-attention mechanism across multi-modal inputs, ViT facilitates joint learning of shared representations, enhancing the model's robustness and utility in real-world scenarios.

For instance, integrating EEG signal analysis with patient demographic or symptomatic data could improve the detection and classification of conditions such as ESES. Furthermore, the pretraining capabilities of ViT on large-scale datasets make it well-suited for transfer learning, reducing the dependency on task-specific data while ensuring high performance.

\section{Conclusion}

This study demonstrates the superior performance of the Vision Transformer model in EEG-based ESES detection, achieving a higher accuracy (97\%) compared to CNNs (94\%) after 80 epochs. ViT's ability to model long-range dependencies, robustness to noise, and compatibility with multi-modal data make it a promising approach for future advancements in medical AI applications. 

The results underscore the potential of transformer-based architectures to outperform traditional CNNs, especially in tasks requiring complex pattern recognition and integration of diverse data types. As demonstrated in this work, ViT not only improves the accuracy of EEG signal analysis but also lays the groundwork for a multi-modal approach to medical diagnostics. This adaptability ensures ViT's relevance in the ever-evolving landscape of machine learning applications in healthcare, paving the way for more inclusive and efficient diagnostic systems.

\newpage

\bibliographystyle{ACM-Reference-Format}
\bibliography{KDD}


\end{document}